\documentclass[a4paper,11pt]{article}
\pdfoutput=1 

\usepackage{jinstpub} 

\let\wt\widetilde

\title{\boldmath Superimposed Electric/Magnetic ``Dipole Moment Comparator'' Lattice Design}

\author{Richard M Talman}
\affiliation{Laboratory for Elementary-Particle Physics,
Cornell University, Ithaca, NY, USA }

\emailAdd{richard.talman@cornell.edu}

\abstract{In contrast to a ``single particle table-top trap'', an essential feature of a storage ring 
``trap'' is that $10^{10}$ or more particles can have their spins aligned in a polarized beam.
This is a nunber of polarized particles large enough for the beam polarization to be detected externally, 
and fed back to permit external control of the beam polarization.  Though the table large enough 
for any such ``storage ring trap'' is quite large, the level of achievable spin control, though 
classical, not quantum mechanical, can be comparable to the control of one or a small number of 
polarized particles in a low energy trap.  

Motivated to investigate time reversal invariance, especially the detection of non-zero electric
dipole moments (EDMs) this 
paper describes the design of a low energy storage ring having the superimposed electric and magnetic 
bending needed to ``freeze'' the spins of polarized beams. For electrons (of either sign) and protons 
the spins can be frozen with all-electric bending but, in general, superimposed electric/magnetic bending is
required.  Since constructive bending superposition in one direction implies destructive superposition in 
the other direction, counter-circulating beams must differ, either in particle type 
or momentum, in order for their orbits to be identical.  

For globally frozen spin operation the bunch polarizations remain constant relative to the momenta, 
for example remaining parallel to the circuating beam momentum vectors. With superimposed electric and magnetic bending, 
the globally frozen spin condition can be met over a continua (specific to particle type) of E/B ratios. When this 
condition is met, the out-of-plane, EDM-induced precession accumulates monitonically, which is obligatory for
producing a measurably large EDM signal.  As Koop has explained, the EDM signal will still accumulate if the
polarization is allowed to ``roll like a wheel'' around a radial axis.
}

\begin{document}
\maketitle
\flushbottom

\section{Introduction}
\label{sec:intro}
This paper discusses the design of a storage ring whose purpose is to detect (T-) time-reversal violation
in the form of non-vanishing electric dipole moments (EDMs). 

As written, the paper is organized much like a \emph{review} article surveying an 
established field in a broad but shallow way.  What makes this ironic is that the paper
can only provide a \emph{preview} of a field that, at the moment, scarcely exists.
With a single exception (a 10\,MeV AGS Analogue Electron Accelerator, conceived of,
designed, built, commissioned, and successfully accomplishing all of its goals, 
before being de-commissioned, all within four or five years in Brookhaven in the 
mid-1950's\cite{Plotkin}\cite{EDM-Challenge}\cite{RT-JDT-AGS-Analogue} no relativistic accelerator 
employing electric bending has ever been built.  

Conventional storage rings have used noticeably large transverse electric fields to 
\emph{separate} counter-circulating beams.  Contrarily, we are concerned with simultaneously 
counter-circulating beams following ``identical'' spatial orbits orbits in rings with
superimposed electric and magnetic bending.  With the bending being constructive in one direction,
and destructive in the other, such a configuration may, superficially, seem to violate T-conservation. 
Not true however; the main motivation for such a perverse pursuit is to search for T-violation.  
Perhaps surprisingly, another goal of the paper is to show how the application of T-invariance can simplify 
the task of designing the storage ring lattice. This includes contemplation of similarities between classical 
and quantum mechanics.

The leading observable effect of a static particle EDM would be an ``out-of-plane'' 
spin precession (orthogonal to ``in-plane'' horizontal spin precession caused by 
whatever magnetic and/or electric fields cause the particle orbit to consist of a sequence
of horizontal circular arcs).  With standard model EDM predictions being much smaller than 
current experimental sensitivities, detection of any particle's non-zero EDM would 
signal discovery of New Physics. 

Currently the proton EDM upper limit (as inferred indirectly by measuring the
Hg atom EDM) is roughly $10^{-24}e \cdot\,$cm\cite{CYR}.
A  ``nominal experimental proton EDM detectability target'' has, by convention, been 
defined to be $10^{-29}e\,\cdot$\,cm.  An EDM of this magnitude could help to account for the 
observed matter/antimatter asymmetry of our universe while, at the same time, being 
plausibly (one or two orders of magnitude) larger than existing standard model 
predictions. This nominal EDM value can also be compared to a general relativistic (GR) 
out-of-plane precession effect, mimicking an EDM of approximately $10^{-28}e\,\cdot$\,cm,
associated with the downward gravitational pull of the earth's magnetic field. 
Depending on storage ring details, this reliably calculable ``background precession'' will 
provide a ``standard candle of convenient magnitude'' calibration of any EDM measurement\cite{CYR}.

\section{Co-magnetometry}
For particles at rest ``co-magnetometry'' in low energy ``table-top particle traps'' has been essential. 
For example, 
Gabrielse\cite{Gabrielse-eEDM} has (with excellent justification) described the measurement of the electron 
magnetic moment (with 13 decimal point accuracy) as ``the standard model's greatest triumph'', based on the 
combination of its measurement to such high accuracy and on its agreement with theory to almost the same accuracy.  

Especially
for the \emph{direct} measurement of EDMs, storage ring technology with beam pairs that can conter-circulate 
simultaneously in a storage ring with superimposed electric and magnetic bending is required.  In this context the term 
``mutual co-magnetometry'' can be used to apply to ``beam type pairings'' for which both beams have frozen spins.  

In the idealized storage ring to be discussed, the electromagnetic
fields are ``cylindrical'' electric ${\bf E}=-E_0{\bf\hat x}r_0/r$ and, 
superimposed, uniform magnetic ${\bf B}=B_0{\bf\hat y}$.
The bend radius is $r_0>0$. Terminology is useful to specify the relative
polarities of electric and magnetic bending:
Cases in which both forces cause bending in the same sense will be called
``constructive'' or ``frugal'';  Cases in which the electric and magnetic
forces subtract will be referred to as ``destructive'' or ``extravagant''.

There is justification for the ``frugal/extravagant'' terminology. 
Electric bending is notoriously weak (compared to magnetic bending) and
iron-free (required to avoid hysteretic effects) magnetic bending is also 
notoriously weak. As a result, an otherwise-satisfactory configuration can be
too ``extravagant'' to be experimentally feasible.

For a particle with spin circulating in a (horizontal) planar magnetic storage ring,
its spin axis precesses around a vertical axis at a rate proportional to the particle's 
anomalous magnetic dipole moment, $G$.  For an ``ideal Dirac particle'' (meaning $G=0$) 
\emph{in a purely magnetic field} the spin precesses at the same rate as the momentum---pointing always 
forward for example.
Conventionally the spin vector's orientation is specified by the in-plane angle $\alpha$ between 
the spin vector ${\bf S}$ and the particle's momentum vector ${\bf p}$ (which is tangential, by definition). 
For such a ``not-anomalous'' particle the spin-tune $Q_M$ 
(defined to be the number of $2\pi$ spin revolutions per particle revolution) 
therefore vanishes, in spite of the fact that, in the laboratory, the spin axis has actually precessed 
by close to $2\pi$ each turn.  

In general, particles are not ideal; the directions of their spin vectors deviate 
at a rate proportional to their anomalous magnetic moments, $G$, and their spin tunes differ from 
zero even in a uniform magnetic field.  Note also, that a laboratory electric field produces a magnetic 
field in the particle rest frame, so a particle in an all-electric storage ring also has, in general, a 
non-vanishing spin tune $Q_E$. Along with $G$ and $Q$, sll of these comments apply equally to the 
polarization vector of an entire bunch of polarized circulating particles.  

By convention, in the BMT-formalism\cite{BMT}\cite{Wolski}, the orientation of the spin vector 
${\bf S'}$ is defined and tracked in the rest frame of the circulating particle, while the 
electric and magnetic field vectors are expressed in the lab. The spin equation of motion 
with angular velocity ${\pmb\Omega}$ is 
\begin{equation}
\frac{d{\bf S'}}{dt} = {\pmb\Omega}\times{\bf S'},
\label{eq:BMT.1} 
\end{equation}
with orbit in the horizontal $(x,z)$ plane assumed, where
\begin{align}
{\pmb\Omega}
 &=
-\frac{q}{\gamma mc}\,
\bigg(\Big(G\gamma\Big)cB_0 + \Big(\big(G - \frac{1}{\gamma^2-1}\big)\gamma\beta^2\Big) \frac{E_0}{\beta}\bigg)\,{\bf\hat y} \notag\\
 &\equiv
-\frac{q}{\gamma mc}\,\bigg((Q_{M})cB_0 + (Q_E)\,E_0/\beta\bigg)\,{\bf\hat y},
\label{eq:BMT.2} 
\end{align}
This equation serves to determine the ``spin tune'', which is defined to
be the variation rate per turn of $\alpha$, as a fraction of 
$2\pi$. Spin tunes in purely electric and purely magnetic rings are given by
\begin{equation}
Q_E = G\gamma - \frac{G+1}{\gamma},
\quad
Q_M = G\gamma,
\label{eq:BendFrac.7}
\end{equation} 
where $\gamma$ is the usual relativistc factor.
Note that the sign of $Q_M$ is the same as the sign of $G$, which is positive for
protons---proton spins precess more rapidly than their momenta in magnetic fields. 
Deuteron spins, with $G$ negative, lag their momenta in 
magnetic fields.  With $G$ positive, $Q_E$ increases from -1 at zero velocity, eventually switching sign
at the ``magic'' velocity where the spins in an all-electric ring are ``globally frozen'' relative 
to the beam direction.  When a particle spin
has precessed through 2$\pi$ in the rest frame it has also completed one full revolution
cycle from a laboratory point of view; so the spin-tune is a frame invariant quantity. 

\section{Superimposed electric and magnetic bending}
\subsection{Circular orbits}
For brevity one can discuss just electrons (including positrons) 
protons($p$), deuterons($d$), tritons($t$), and helions($h$), or even just $p$ 
and $d$, based on the consideration that most of the apparatus, and all of the 
technology, needed for their EDM  measurement is presently available at COSY
laboratory in Juelich, Germany.

The circulation direction of a so-called 
``master beam'' (of whatever charge $q_1$) is assumed to be CW or, equivalently,
$p_1>0$. A secondary beam charge $q_2$ is allowed to have either 
sign, and either CW or CCW circulation direction.

Ideally both beam polarizations would be frozen ``globally'' (meaning spin tune $Q_S$ is zero and the angle $\alpha$ 
between polarization vector and momentum is constant everywhere around the ring).  (Somewhat weaker) ``doubly-frozen'' 
can (and will) be taken to mean that a ``primary beam'' locked to $Q_S=0$, circulates concurrently with a ``secondary'' beam that 
is ``pseudo-frozen'', meaning the spin tune is locked to an unambiguous, exact, rational fraction.  Only if this
rational fraction is zero, would the terminology ``doubly-magic'' be legitimate.

These pairings are expected to make direct EDM difference measurements of unprecedented precision possible.  For any 
arbitrary pairing of particle types ($(p,d),\ (p,e-),\ (\mu,e+),\ (d,h),\ (p,t)$, etc.) continua of such doubly-frozen 
pairings are guaranteed.

A design particle has mass $m>0$ and charge $qe$, with electron charge 
$e>0$ and $q=\pm 1$ (or some other integer). These values produce circular 
motion with radius $r_0>0$, and velocity ${\bf v}=v{\bf\hat z}$, where the motion
is CW (clockwise) for $v>0$ or CCW for $v<0$. With $0<\theta<2\pi$ being 
the cylindrical particle position coordinate around the ring, the angular 
velocity is $d\theta/dt=v/r_0$. 

(In MKS units) $qeE_0$ and $qe\beta c B_0$ are commensurate forces, 
with the magnetic force relatively weakened by a factor $\beta=v/c$ 
because the magnetic Lorentz force is $qe{\bf v}\times{\bf B}$. 
By convention $e$ is the absolute value of the electron charge; where it
appears explicitly, usually as a denominator factor, its purpose in 
MKS formulas is to allow energy factors to be evaluated as electron volts (eV)
in formulas for which the MKS unit of energy is the joule. 
Newton's formula for radius $r_0$ circular motion, expressed in terms of 
momentum and velocity (rather than just velocity, in order to be relativistically valid)
can be expressed using the total force per unit charge in the form
\begin{equation}
\beta\frac{pc}{e} = \Big(E_0 + c\beta B_0\Big)\,qr_0,
\label{eq:CounterCirc.1} 
\end{equation}
Coming from the cross-product Lorentz magnetic force, the factor $q\beta cB_0$
is negative for backward-traveling orbits because the $\beta$ factor 
is negative.

A ``master'' or primary beam travels in the ``forward'', CW direction. 
For the secondary beam, the $\beta$ factor can have either sign.
For $q=1$ and $E_0=0$, formula~(\ref{eq:CounterCirc.1}) reduces to a standard 
accelerator physics ``cB-rho=pc/e'' formula.  For $E_0\ne 0$ the formula 
incorporates the relative ``bending effectiveness'' of $E_0/\beta$ 
compared to $cB_0$.  As well as fixing the bend radius $r_0$,
this fixes the magnitudes of the electric and magnetic bend field values 
$E_0$ and $B_0$. To begin, we assume the parameters of a frozen spin ``master'',
charge $qe$, particle beam have already been established, including the signs
of the electric and magnetic fields consistent with $\beta_1>0$ and $p_1>0$.  
In general, beams can be traveling either CW or CCW.  For a CCW beam both $p$ and 
$\beta$ have reversed signs, with the effect that the electric force is unchanged, but the 
magnetic force is reversed. The $\beta$ velocity factor can be expressed as
\begin{equation}
\beta = \frac{pc/e}{\sqrt{(pc/e)^2 + (mc^2/e)^2}}.
\label{eq:CounterCirc.2} 
\end{equation}
Eq.~(\ref{eq:CounterCirc.1}) becomes
\begin{equation}
\frac{pc}{e} = \Big(\frac{E_0\sqrt{(pc/e)^2 + (mc^2/e)^2}}{pc/e} + cB_0\Big)qr_0.
\label{eq:CounterCirc.3} 
\end{equation}
Cross-multiplying the denominator factor produces
\begin{equation}
\Big(\frac{pc}{e}\Big)^2 = qE_0r_0\sqrt{(pc/e)^2 + (mc^2/e)^2} + qcB_0r_0\frac{pc}{e}.
\label{eq:CounterCirc.4} 
\end{equation}
To simplify the formulas we make some replacements and alterations, 
starting with
\begin{equation}
pc/e \rightarrow p,
\quad\hbox{and}\quad
m c^2/e\rightarrow m, 
\label{eq:Alterations.1}
\end{equation}
The mass parameter $m$ will be replaced later
by, $m_p$, $m_d$, $m_{\rm tritium}$, $m_e$, etc., as approppriate
for the particular particle types, proton, deuteron, triton, electron, helion, etc..
These changes amount to setting $c=1$ and switching the energy units from joules to electron volts. 
The number of ring and beam parameters can be reduced by forming the combinations 
\begin{equation}
\mathcal{E} = qE_0r_0,
\quad\hbox{and}\quad
\mathcal{B} = qcB_0r_0.
\label{eq:Alterations.2}
\end{equation}
After these changes, the closed orbit equation has become  
\begin{equation}
p_m^4 -2\mathcal{B}p_m^3 + (\mathcal{B}^2-\mathcal{E}^2)p_m^2 - \mathcal{E}^2m^2=0,
\label{eq:AbbrevFieldStrengths.3}
\end{equation}
an equation to be solved for either CW and CCW orbits.  The absence of a term linear in $p_m$
suggests the restoration, using Eq.~(\ref{eq:Alterations.2}), of the explicit form of 
$\mathcal{B}$ in the coefficient of the $p_m^3$ term to produce;
\begin{equation}
p_m^4 - 2cB_0(qr_0)p_m^3 + (\mathcal{B}^2-\mathcal{E}^2)p_m^2 - \mathcal{E}^2m^2 = 0,
\label{eq:AbbrevFieldStrengths.3-rev} 
\end{equation}
The product factor $(qr_0)$ can be altered arbitrarily without influencing any conclusions.
This and other properties can be confirmed by pure reasoning, based on the structure of the equation,
or by explicit partially-numerical factorization of the left hand side.

These considerations have removed some, but not all of the sign ambiguities introduced by the quadratic 
substitutions used in the derivation of Eq.~(\ref{eq:AbbrevFieldStrengths.3-rev}). 
The electric field can still be reversed without altering the set of solutions of the equation. 
Note that this change cannot be compensated by switching the sign of $q$, which also reverses the 
magnetic bending.  The most significant experimental implication is that it is not only positrons,
but also electrons, that can have orbits identical to (necessarily positive in practice) baryons.

We can contemplate allowing the signs of $E_0$ or $B_0$ to be reversed for experimental purposes, such as interchanging CW and
CCW beams, or replacing positrons by electrons, but only if this can be done 
with sufficiently high reproducabilty.  \emph{Demonstrating this capability (by promising spin tune measurability
with frequency domain precision) is an important ingredient of this paper.}

Fractional bending coefficients $\eta_E$ and $\eta_m$ can be defined by
\begin{equation}
\eta_E = \frac{qr_0}{pc/e}\,\frac{E_0}{\beta},\ 
\eta_M = \frac{qr_0}{pc/e}\,cB_0,
\label{eq:BendFrac.2}
\end{equation}
neither of which is necessarily positive.  These fractional
bending fractions satisfy
\begin{equation}
\eta_E + \eta_M = 1\quad\hbox{and}\quad
\frac{\eta_E}{\eta_M} = \frac{E_0/\beta}{cB_0}.
\label{eq:BendFrac.2p}
\end{equation}
The ``potencies'' of magnetic and electric bending are in the ratio $cB_0/(E_0/\beta)$ because  the electric
field is stronger than the magnetic by the factor $1/\beta$ as regards bending charge $q$ onto an orbit with 
the given radius of curvature $r_0$. The curious parenthetic arrangement of Eq.~(\ref{eq:BMT.2}) is intended 
to aid in the demonstration that, when expressed in term of spin tunes, the  ``potencies'' of magnetic and electrically 
induced MDM precessions are in the same ratio as the bending potencies. 

\subsection{Frozen spins}
The combined field spin tune can be expressed in terms of the fractional precession coefficients;
\begin{equation}
Q_S = \eta_EQ_E + \eta_MQ_M.
\label{eq:BendFrac.6} 
\end{equation}
Superimposed electric and magnetic bending permits beam spins to be frozen 
``frugally''; i.e. with a ring smaller than would be required for all-electric 
bending; for spin tune $Q_S$ to vanish requires
\begin{equation}
Q_S = \eta_{_E}Q_E + (1-\eta_{_E})Q_M = 0.
\label{eq:SpinPrecess.5m}
\end{equation}
Solving for $\eta_{_E}$ and  $\eta_{_M}$, 
\begin{equation}
\eta_{_E} = \frac{G\gamma^2}{G+1}, \quad
\eta_{_M} = \frac{1+G(1-\gamma^2)}{G+1} = \frac{1-G\beta^2\gamma^2}{G+1}.
\label{eq:SpinPrecess.5}
\end{equation}
For example, with proton anomalous magnetic moment $G_p=1.7928474$, trying $\gamma=1.25$, 
we obtain $\eta_{_E} = 1.000$ which agrees with the known proton 
233\,Mev kinetic energy value in an all-electric ring. 
For protons in the non-relativistic limit, 
$\gamma\approx1$ and $\eta^{\rm NR}_E \approx2/3$.

The electric/magnetic field ratio for the primary beam to be frozen is
\begin{equation}
\frac{\eta_E}{\eta_M} = 
\frac{E_0/\beta}{cB_0} = \frac{G_1\gamma_1^2}{1-G_1\beta_1^2\gamma_1^2}.
\label{eq:SpinPrecess.5pp}
\end{equation}
For given $\beta_1$, along with this equation and the required bend radius $r_0$, this fixes the 
electric and magnetic fields to the unique values that globally freeze the primary beam spins.
With $1\rightarrow2$ subscript replacement, the same frozen beam formulas apply to the secondary 
beam; note, though, that the $\beta$ factor has opposite sign. To be ``doubly-magic'' both beams must satisfy this relation.

\section{Symplecticity-assisted lattice design}
\subsection{Superimposed electric/magnetic lattice complications}
The fundamental complication of an electric ring, as contrasted 
with a magnetic ring, is the non-constancy of particle speed\cite{RT-RAST}.
A fast/slow separation into betatron and synchrotron amplitudes 
has become fundamental to the conventional Courant-Snyder (CS) 
magnetic ring formalism. For CS, since the mechanical energy varies
only in RF cavities, the $\gamma$ factor is invariant in the rest of the ring, and one is 
accustomed to treating $\gamma$ as 
constant for times short compared to the synchrotron period. 
Only to the extent the betatron parameters are independent of 
total particle energy, can the betatron and synchrotron motions 
be directly superimposed.

By contrast, in an electric lattice the mechanical energy (as quantified by $\gamma$) 
varies on the same time scale as the transverse $x$ and $y$ amplitudes.
On the other hand, the slow change, only in RF cavities, 
of the total energy $\mathcal{E}=\gamma mc^2+eV(r)$, which includes
also the potential energy $eV(r)$, makes a similar fast/slow 
separation possible. 

To most closely mimic the fast/slow superposition of betatron
and synchrotron oscillations in an electric ring,
and to continue to regard $\gamma$ as the fundamental
``energy-like'' parameter, requires us to evaluate 
$\gamma$ only in regions of zero electric potential,
which is to say, not in RF cavities, and not in electric
bending elements---in other words, only in field free
drift regions. This leads to a curious, but entirely
manageable, representation in which the particle orbits
are modeled exactly only in drift regions, though most of their 
time is spent inside bend elements where $\gamma$ is variable, and  
little time in short drift regions (where $\gamma$ is constant). 

The reason this approach is fully satisfactory
is that the drift regions are fairly closely spaced, and more or 
less uniformly distributed around the ring. Knowing the lattice
functions exactly in these regions is operationally almost
as satisfactory as knowing them everywhere.
With these qualifications, one can still rely on the approximate 
representation of individual particle motions
as a superposition of fast betatron and slow synchrotron motions.

\subsection{Transfer matrix evolution}
It is important to notice, in subsequent sections, that there is no mention of the source of 
bending and focusing.  Irrespective of the electric/magnetic character of the elements
in an accelerator, particle orbits (which, for simplicity, we take to be 
executing only small amplitude vertical betatron oscillation, are 
focused by ring lattice elements of focusing strength $K(s)$, where $s$ is a tangential
coordinate along the design (or central) orbit such that the trajectory
satisfies the ``focusing differential equation''
\begin{equation}
\frac{d^2y}{ds^2} = K(s)y.
\label{eq:Beams.1}
\end{equation}
The sign of $K$, like that of a Hooke's law force, is negative for ``restoring''.
(In practice, one way or another, the focusing is always ``alternating gradient'' (AG),
so, locally, the sign is as likely to be positive as negative---and of opposite sign
for horizontal betatron oscillations.)

The dependence of $K(s)$ on $s$ permits the description of 
systems in which the focusing strength varies along the orbit. In particular, $K(s)=0$
describes ``drift spaces'' in which case Eq.~(\ref{eq:Beams.1}) is
trivially solvable, and yields the obvious result that particles in free space travel in 
straight lines.

\subsection{Design methodology}
(Deferred until the methodology used in its design has been described) a layout of the full ring,
(to be referred to here as ``BSM) is 
shown in FIG~\ref{fig:PTR-layout-0p80-COSY-4PAX}. The ring has 
super-periodicity $n_c=4$. Optical elements for one super-period are shown on the left. Since each quadrant 
is forward/backward symmetric, it is sufficient to design, and display, just 
one eighth of the ring. $\beta_x$ and $\beta_y$ are plotted against element indices (ordinal numbers, 
starting at 1) in FIG~\ref{fig:labeled-beta-functions-NOMINAL-map-index}. (Barely visible) grid lines mark 
the boundaries between adjacent elements. 

The lattice design has been performed using a program, MAPLE-BSM, that exploits the algebraic 
(as contrasted with numerical) capabilities of typical lattice analysis programs.  This \emph{design code}
is based on Wollnik transfer matrix elements\cite{Wollnik}, which implicitly describe orbit evolution between
points of zero electric potential energy.

One sees that the ring is very simple since the element index increases by 1 
from element to element, and the figure is mirror symmetric about map index 10. All element names, including
drift lengths, are shown, drifts above, powered elements below.

For brevity, we describe only vertical motion, and describe the evolution of vertical phase 
space coordinates ${\bf y}=(y,y')^T$, a two component column vector, by transfer matrix multiplication;
\begin{equation}
{\bf y}_1 = {\bf M_{10}}{\bf y}_0 .
\label{eq:M01}
\end{equation}
To obtain the once-around transfer matrix at location~1, one starts by calculating ${\bf M_{10}}$,
the transfer matrix from map index 0 to 1; note that the matrix indices are attached ``backwards'',
\emph{not} in increasing index order. Exploiting symplecticity, from ${\bf M_{10}}$ one can obtain 
${\bf M^{-1}_{10}}$ algebraically; (i.e. analytically, not numerically.)\footnote{``Algebraic'' design 
implies that, in principle, an entire lattice design can be performed in closed form. In practice this 
would be impossible, since there are far too many independent parameters. The combinatorics of handling 
a lrge number of independent arguments could overwhelm even the most powerful computer program in 
the most powerful computer.  But, with care in introducing free parameters, all design procedures, such as 
inverting matrices and solving constraint equations, can be handled in closed form---with
numerical values produced only for output convenience.} The algebraic relation is
\begin{equation}
{\bf M}^{-1} = -{\bf S}{\bf M}^T{\bf S},
\quad\hbox{where}\quad {\bf S} = \begin{pmatrix} 0 & -1 \\ 1 & 0  \end{pmatrix}
\label{eq:M01p}
\end{equation}
for $2\times2$ matrices and ${\bf S}$ is replicated along the diagonal for higher dimensions.
One can proceed to find ${\bf M_{21}}$ and ${\bf M^{-1}_{21}}$ and so on, 
in the same way.  Propagation from 0 to 2 is given by 
\begin{equation}
{\bf M_{20}} = {\bf M_{21}} {\bf M_{10}},
\label{eq:M02}
\end{equation}
and so on. Iterating these calculations, one next describes motion through just one
of the $n_C$ super-periods.  Then, by just $n_C$ more matrix multiplications
one can find ${\bf M_{00}}$, the ``once-around transfer matrix'' at the origin.
The once-around transfer matrix at location~1, ${\bf M_{11}}$, is then given by
\begin{equation}
{\bf M_{11}} = {\bf M_{10}} {\bf M_{00}} {\bf M^{-1}_{10}},
\label{eq:once-around}
\end{equation}
One notes that, whereas the orbit coordinates $(y,y')$ evolve by direct transformation, 
the lattice parameters evolve by similarity transformation. This duality resembles the
Schrodinger/Heisenberg complementary ``pictures'' in quantum mechanics.

The Twiss parameters, $\alpha$, $\beta$ and $\mu$ can be solved for at every location using 
the four equations implied, element by element, by the equation
\begin{equation}
{\bf M}(s,s+\mathcal{C})
 =  
\begin{pmatrix} 
      \cos\mu + \alpha(s)\sin\mu                            &    \beta(s)\sin\mu          \\
  \frac{-(1+\alpha^2(s))\sin\mu + \alpha\cos\mu}{\beta(s)}  &  \cos\mu - \alpha(s)\sin\mu
\end{pmatrix}
\label{eq:Beams.10}
\end{equation}
where $\alpha=-\beta'/2$; $\mu\equiv2\pi Q$ is the ``ring phase advance''; and $Q$
is the ``ring tune''. Equally important, in a ring  with super-periodicity $n_C$,
the same formula is valid, with $\mu\rightarrow\mu/n_C$ and 
$\mathcal{C}\rightarrow\mathcal{C}/n_C$

Proceeding inductively, 
one obtains once-around transfer matrices and Twiss parameters at every element interface. 
They are plotted in FIG~\ref{fig:labeled-beta-functions-NOMINAL-map-index}. The points are joined 
by straight lines. This is mildly misleading since, with $\alpha(s)=-\beta'(s)/2$ being continuous, 
the ``kinks'' visible in $\beta(s)$ are artifical, and need to be ``rounded off''
mentally. Also, plotted against element index, the beta function shape is distorted from what one is
accustomed to seeing.  This is rectified, in the subsequent plots, by plotting $\beta(s)$---but the kinks, 
caused by straight line interpolation, remain.

\subsection{Lattice \emph{design} and lattice \emph{analysis} contrasted}
There are  many lattice simulation programs, SYNCH, MAD, MADX, TEAPOT, PTR, B-MAD, ELEGANT,
to name just a few. All of these are primarily ``lattice analysis programs''---a term to be
defined (unconventionally) below. \emph{Starting from a sequential list of design elements:} bending elements, quadrupoles, 
sextupoles, RF cavities, beam position monitors, along with their lengths, strengths, and all other relevant
parameters, these programs support lattice \emph{analysis}.  
As well as providing long term particle position tracking (and spin orientations if
necessary) such programs provide for setting the strengths of all the powered elements
to flatten the orbit, set the tunes, adjust the focusing properties, and so on.

Commonly the lattice description inputs to such programs can be idealized, in the
sense that many elements have identical parameters and identical powering; this feature is
supported by allowing the parameters to be algebraic, rather than numeric. Eventually though, 
to allow for their not quite identical properties, some or, in general, all, of their parameters
have to be ``fully-instantiated'', meaning numerical rather than algebraic. Typically the fitting
algorithms mentioned in the previous paragraph are entirely numerical, though with methods for
grouping elements into ``families of elements'' whose strengths are constrained to scale 
proportionally. 

All these features can be provided by computer languages such as Fortran, C, and C++, Python, etc.
In the terminology I have been employing all these programs are \emph{analysis} programs employing numerical
algorithms.  What they \emph{are not}, is ``design programs'' capable of taking advantage of powerful 
symbolic (i.e. algebraic) formula manipulation, and equation solvers, such as MAPLE and MATHEMATICA. 
Familiar, myself, only with MAPLE, I assume that the capabilities of these two computational languages
(and perhaps others) are more or less equivalent. The single most essential ``solving mechanism'' requirement 
for a design code is the abilty to invert or diagonalize matrices.  

\subsection{Resemblance of lattice design to quantum mechanics}
With $y\rightarrow\psi$, and $K(s)\rightarrow 2m(E- U(s)/\hbar^2$,
one notes that Eq.~(\ref{eq:Beams.1}) becomes the
Schr\"odinger equation satisfied by a stationary plane wave as the wave function 
for a particle of energy $E$ in a potential $U(s)$. With all storage ring beam particles 
being paraxial and all traveling at nearly the same constant speed $v$, their longitudinal 
components advance in time as $s=vt$.  The further replacement $\psi\rightarrow s-vt$
produces a travelng wave not unlike a betatron oscillation
This suggests some kind of duality between waves and particles. Acually Newton was 
aware of this duality 400 years ago, both experimentally and theoretically. And
wave/particle duality has been around ever since.  

Returning to our Eqs.~(\ref{eq:Beams.1}) and (\ref{eq:once-around}) we have another
kind of duality. In a ``Schr\"odinger-like picture'', from initial conditions 
$(y_0,y'_0)$ it is natural to visualize the evolution of phase space coordinates 
$\big(y(t),y'(t)\big)$ with increasing $t$ (or $s$) as solving Eq.~(\ref{eq:Beams.1})
directly.  (\emph{As an aside, it should be noted that the function $K(s)$ is assumed to be
known, which means the ring lattice has already been \emph{designed}.})
But, in a Heisenberg-like picture, one can visualize initial conditions  
$(y_0,y'_0)$ as parameterizing a fixed state in which $\beta(s)$ is a 
particle or beam dynamic variable evolving according to
Eq.~(\ref{eq:once-around}) or by equivalent matrix operator. \emph{Accelerator physicists are
ambivalent as to whether beta functions are properties of a beam or of a lattice.}

This duality may seem to be of only academic interest. Eq.~(\ref{eq:Beams.1}) is 
linear and simple, and Eq.~(\ref{eq:once-around}) is nonlinear and complicated. On the
other hand, for linearized amplitude $a$ motion, Eq.~(\ref{eq:once-around}) does \emph{provide a solution in
conveniently parameterized form; }
\begin{equation}
y(s) = a\sqrt{\beta(s)}\,\cos(\mu-\mu_0).
\label{eq:Beams.5}
\end{equation}
Following sections provide less superficial distinctions between the approaches.

The case being made is that it is
sensible to \emph{design} a new lattice in a Heisenberg-like picture even if one
is intending to \emph{analyse} its performance primarily using a Schr\"odinger-like picture.
Justification for this will be expanded below but, briefly, the design process is
inherently nonlinear, with a vast number of initial parameters needing to be fixed, 
while the analsis process is, in lowest approximation, linear, with nonlinearity
entering only perturbatively. 

Continuing to dwell on classical (CM) and quantum (QM) duality,  
one notes that both disciplines require all physically measureable quantities to be real,
not complex, numbers.  In QM, even though wave functions are allowed to be complex,
physically measureable quantities need to be represented by the (real)
eigenvalues of Hermitean operators, even though Hermitean matrices or their infinite 
dimensional generalizations, typically have complex components.

In accelerator CM the wave functions are physically measureable particle positions
and momenta, all of which need to be real. CM transfer matrices are by no
means Hermitean, as can be confirmed from any transfer matrix introduced so far. 
Furthermore the elements of CM transfer matrices must also be measureable quantities
that are necessarily real.  In general, therefore, the eigenvalues of CM transfer
matrices are complex.  Clearly, then, even when related QM operator matrices and 
CM transfer matrices have identical dimensionality, they cannot, in any sense, play
analogous roles.  There is no ``Hermitean-like'' trick in CM guaranteing that a derived
$\beta(s)$ fuction meets the necessary condition of being real and positive. 
This has to be handled in the ``old-fashioned way''---when solving a quadratic equation, 
of selecting only real roots. Any lattice \emph{designer} knows that, 
in practice, at first cut, the value of $\cos\mu$, appearing in Eq.~(\ref{eq:Beams.10}), 
rarely lies in the rage $-1< \cos\mu<1$---as it must for ring stability---without 
careful fiddling of ring lattice parameters.
  
A feature shared by CM and QM is that they are both Hamiltonian. Though this
is an exact requirement in QM, it is only an approximate requirement in CM. 
The reason it is only approximate is that classical mechanical systems (the only
kind of mechanisms we have at our disposal) are invariably ``lossy''---the Q-value
(quality factor) of the highest quality resonators, though very large compared to 1,
are small compared to infiniity. So there is always time reversal violation at
some level in practical classical mechanics. 

The nearest exception to this general statement about classical mechanics is the sort 
of low energy hadron accelerator under discussion in the present paper. With synchrotron 
radiation virtually absent, protons or deuterons can circulate losslessly for days. 
What guarantees this in the Courant-Snyder accelerator formalism is that the CM transfer 
matrices are ``symplectic'', a term synonymous, in general, with ``Hamiltonian''. 

It is well known to accelerator physicists that transfer matrices have to be symplectic.
Such physicists, occasionally, and disreputably, counter (erroneously-) calculated damping of the 
Courant-Snyder invariant---the quantiy that symplecticity guarantees conserved---artificially 
``re-symplectify'' the formalism in use, even at the possible cost of violating energy 
conservation.  This issue is too esoteric to be pursued at the level of the 
present paper.

In passing, it can also be mentioned that, in classical mechanics, symplectic transformations preserve 
Poisson brackets\cite{RT-GM}. It is also well known that the bridge between CM
and QM consists, primarily, of the replacement of Poisson brackets of classical quantities by
the commutators of their QM replacements.

Less well known is that it is trivially easy to invert a symplectic matrix algebraically---it need 
not be done numerically. It is this feature which, I hope, is sufficiently important to justify 
such a lengthy and abstract build-up as has been given to this point. It is the exploitation of this
feature that enables flexible lattice design features to be coded easily into brief MAPLE
or MATHEMATICA programs.  

My lattice design program MAPLE-BSM exploits capabilities these high level computer languages 
have, that lower level languages do not have, to design a storage ring lattice that can store simultaneously 
counter-circulating frozen spin beams in a ring with superimposed electric and magnetic bending. 
BSM is an acronym for ``Belt and Suspenders, Mutable'', with the implied meaning that the ring focusing
is redundantly provided by (very weak) alternating gradient focusing provided by electrode shaping,
indicated by $m<0$, $m>0$ labels in FIG~\ref{fig:PTR-layout-0p80-COSY-4PAX}, 
along with separated function quadrupoles labelled Qf, Qd, Qir1, and Qir2.
Features of the program have been mentioned but details of the program are documented separately.  

\section{BSM: a Belt and Suspenders, Mutable, symmetry-violation sensitive lattice}
\subsection{Lattice properties}
\begin{figure}[h]
\centering
\includegraphics[scale=0.5]{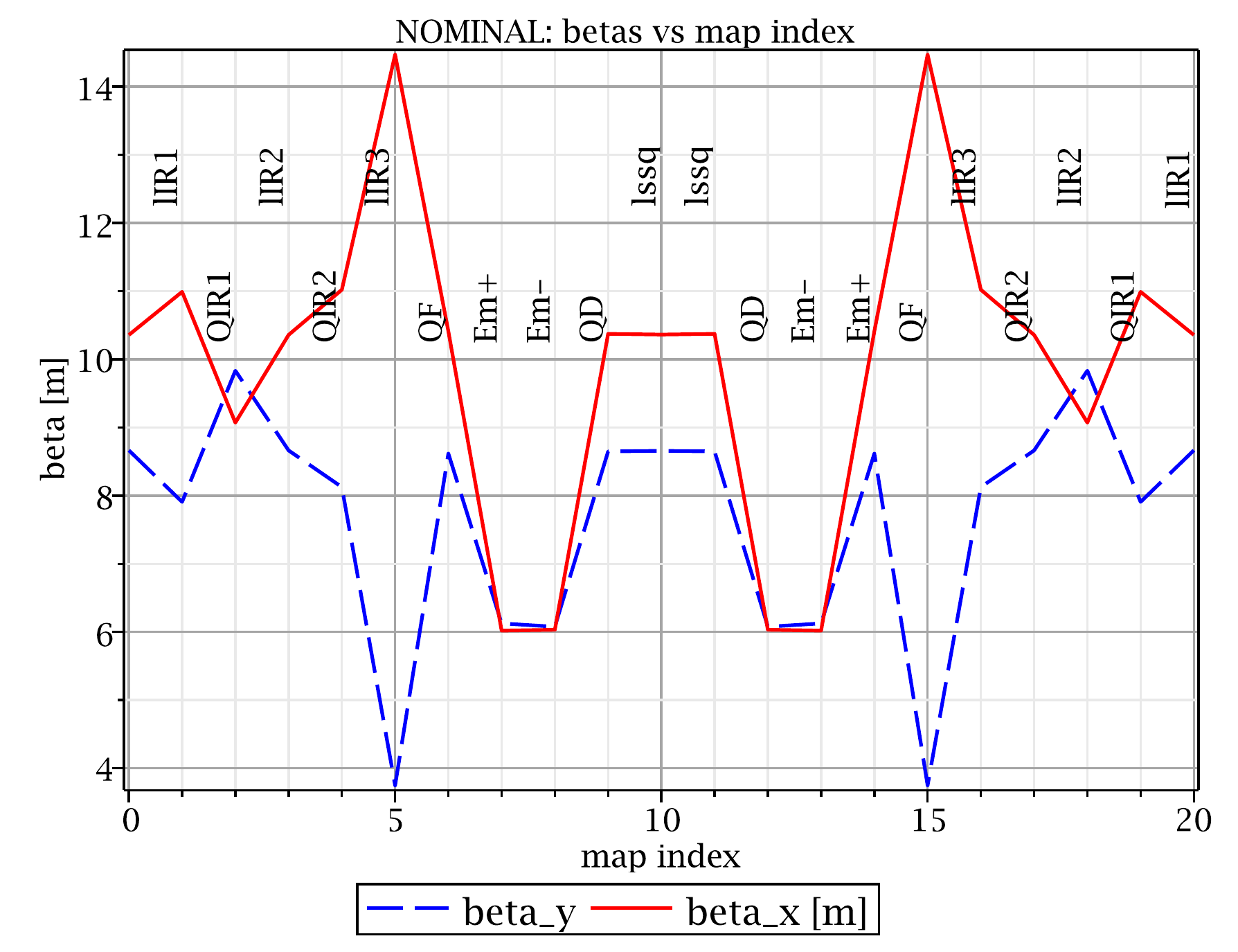}
\caption{\label{fig:labeled-beta-functions-NOMINAL-map-index}$\beta_x$ and $\beta_y$ are 
shown plotted against element indices, for one quadrant of the full ring. With super-periodiciy of
4, the other three quadrants are identical.  In this plot (only) the length scale is arbitrary
(within reason). In all subsequent lattice function plots the horizontal axis is $s$.}
\end{figure}

\begin{figure}[hbt]
\centering
\includegraphics[scale=0.26]{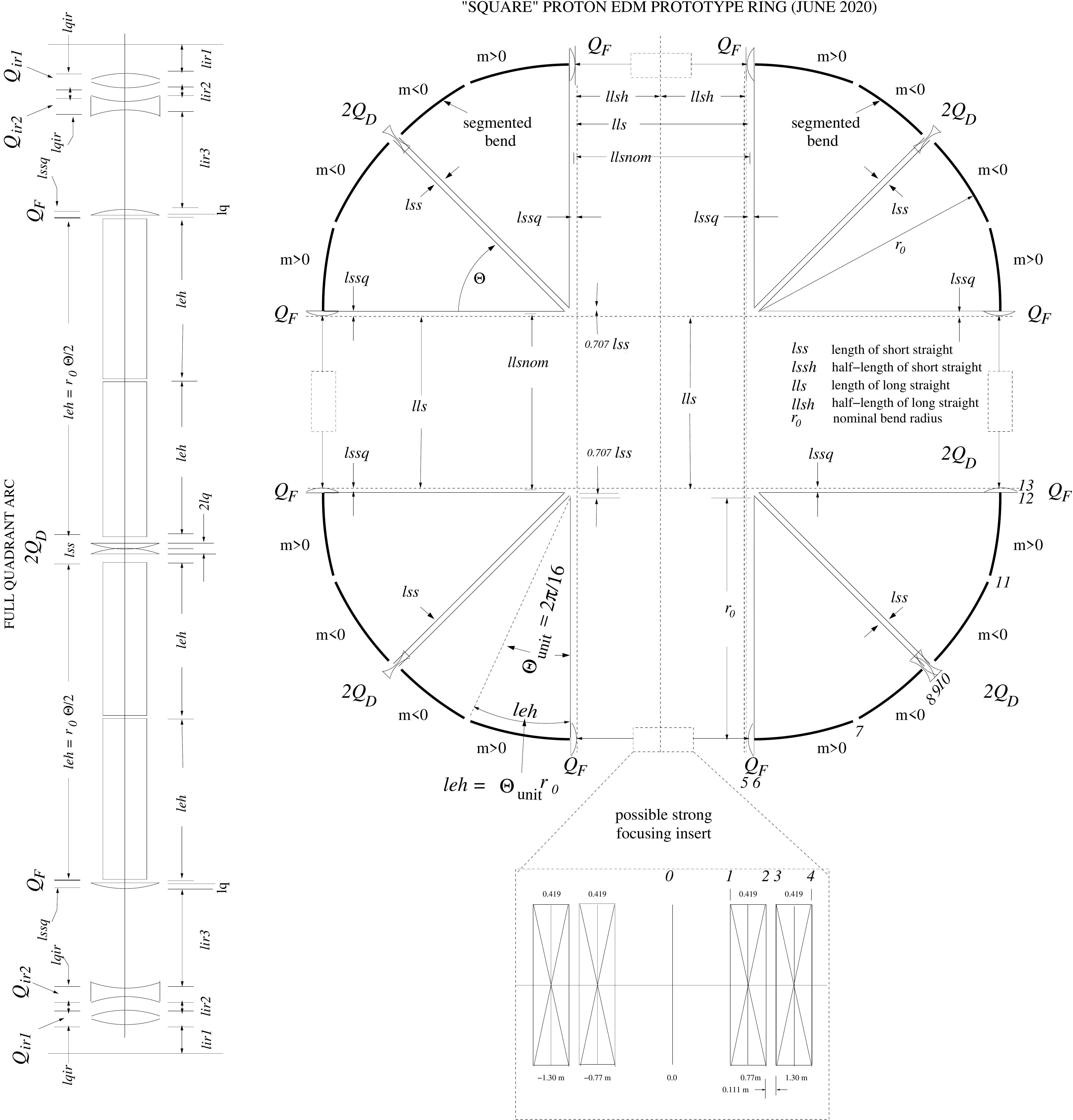}
\caption{\label{fig:PTR-layout-0p80-COSY-4PAX}Lattice layouts for a proposed ``EDM prototype'' 
storage ring: single quadrant (left) and full ring geometry (right). (For pictorial 
convenience quadrupole symbols represent neither actual quadrupole lengths nor fine-grained 
locations, and may also subsume sextupoles not shown.) A doublet-pair present in every 
straight secion is broken out only in the lower (south) straight section.  
Increased quadrupole doublet strengths in all four long straights converts the lattice to 
"strong-focusing" (though not \emph{very} strong by modern standards.)  In any case, 
the total accumulated drift length is not enough for the ring to operate ``below transition''.  
When scaling up to the eventual, full energy, all-electric ring, from four-fold to sixteen-fold 
symmetry, with drift lengths and bend lengths preserved (but bend angles four times less) the 
total circumference is to be 500\,m or greater and operation will be well below transition. }
\end{figure}

\begin{figure}[h]
\centering
\includegraphics[scale=0.45]{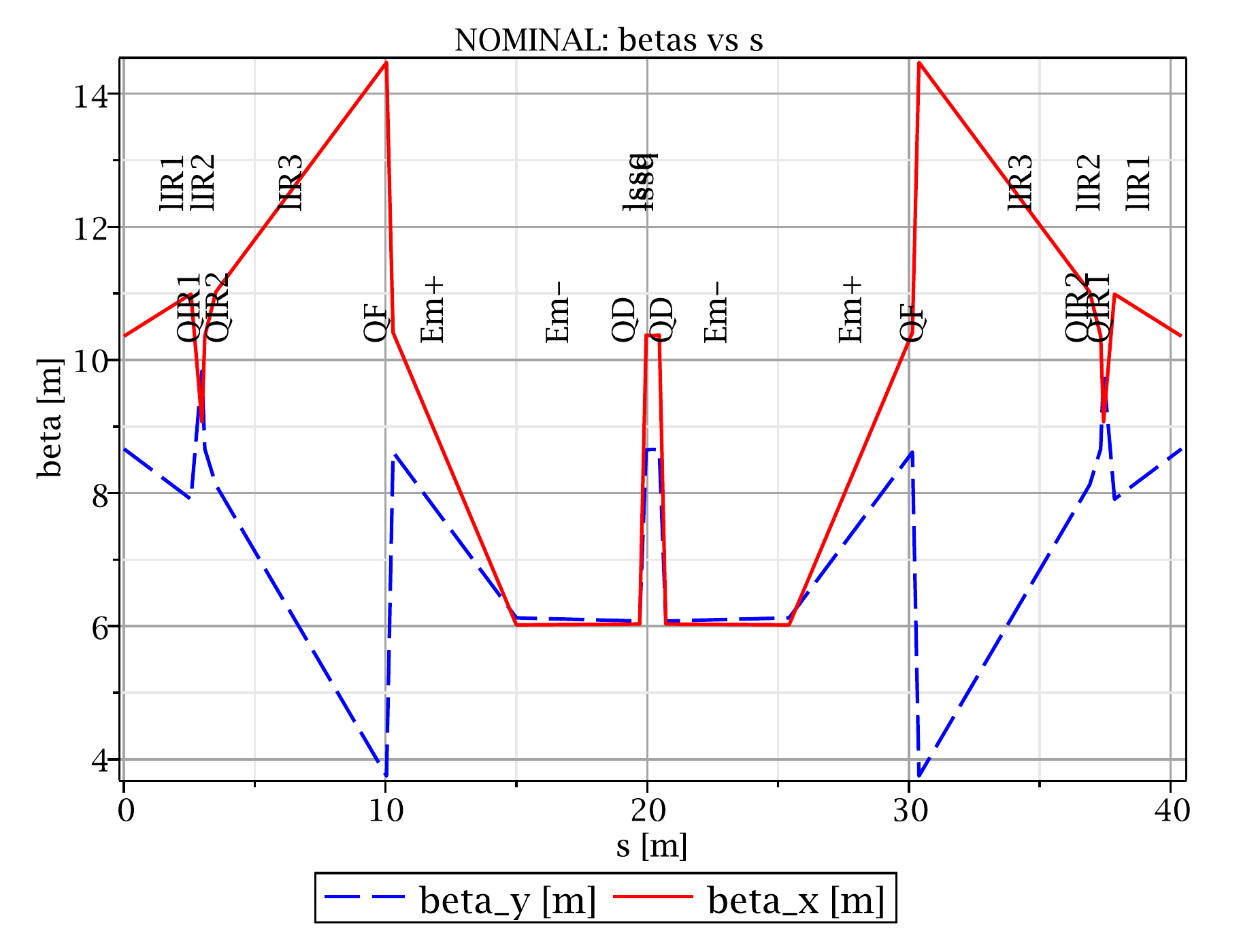}
\includegraphics[scale=0.45]{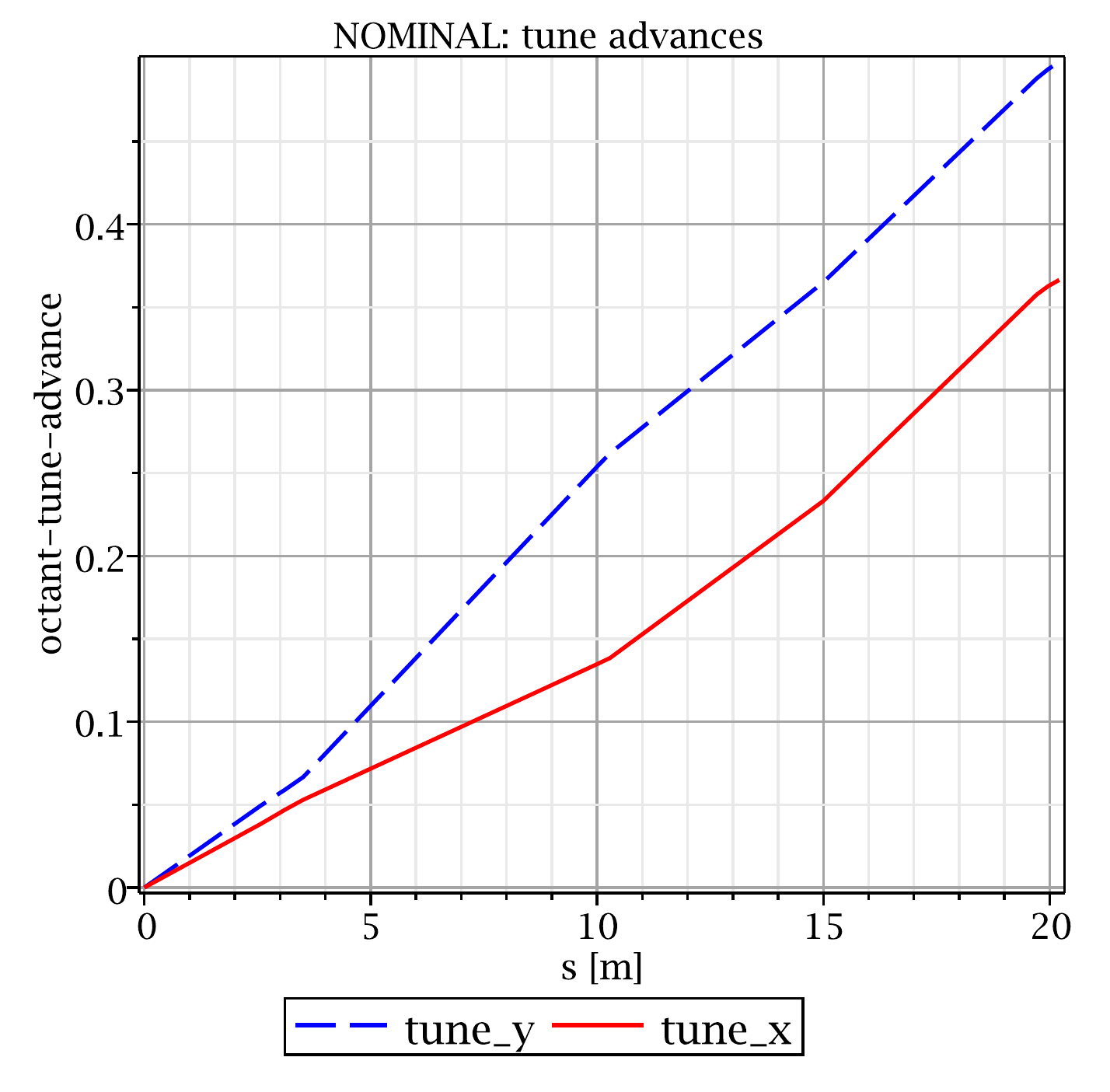}
\caption{\label{fig:labeled-beta-functions-NOMINAL-s}The data for the figure on the left is the same as 
in the previous FIG~\ref{fig:labeled-beta-functions-NOMINAL-map-index} but, with horizontal axis registering
accumulating tangential coordinate $s$. Tune advances for one quadrant are plotted against $s$ on the right.  
Since each quadrant is mirror-symmetric it is sufficient to display just one octant (and confusing to display
the accumulating tune advances). The full-ring tunes are roughly 3.5 in both planes. }
\end{figure}

\begin{figure}[h]
\centering
\includegraphics[scale=0.52]{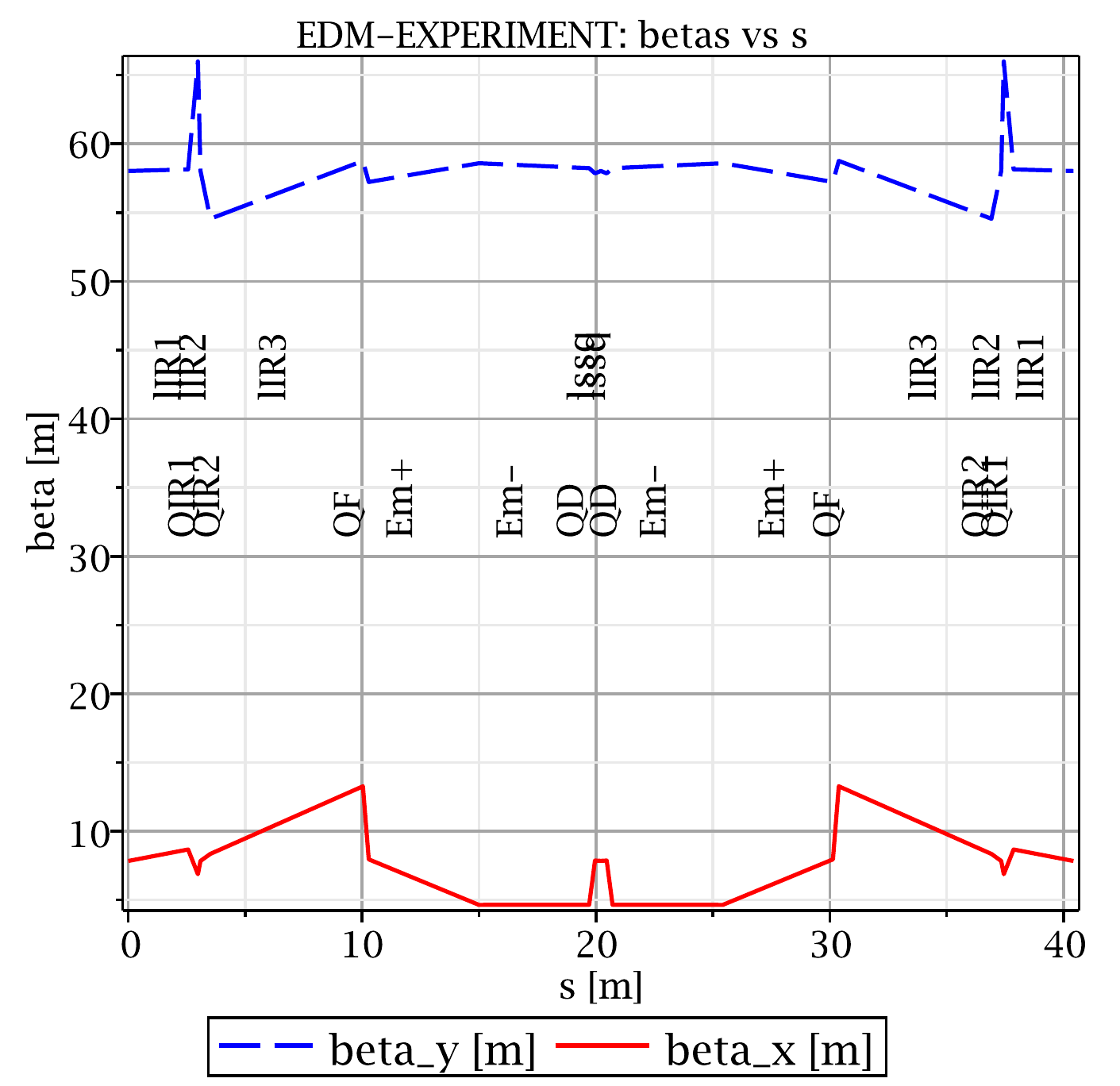}
\includegraphics[scale=0.52]{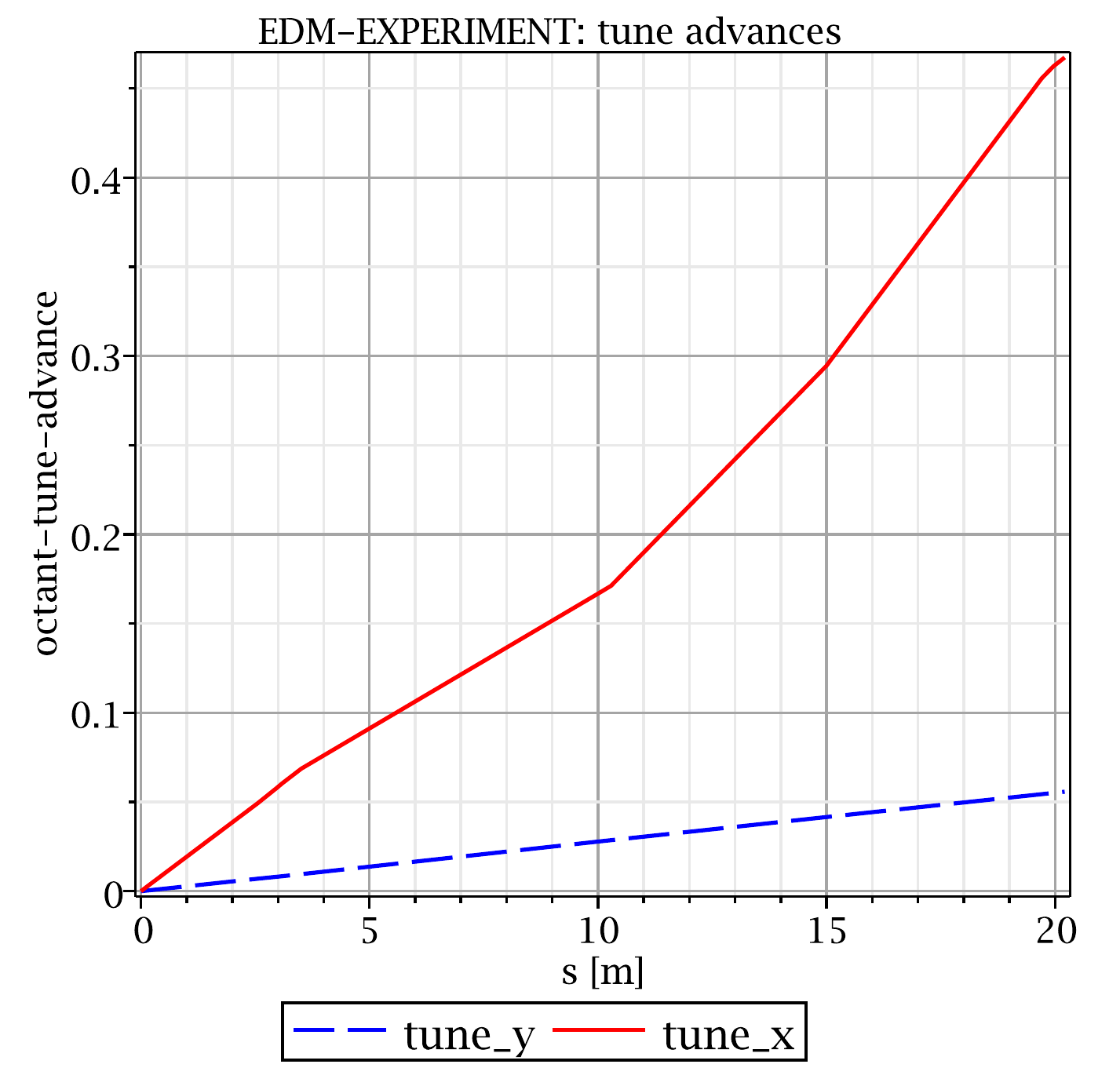}
\caption{\label{fig:labeled-beta-functions-EDM-EXPERIMENT-s}This plot provides the same information as 
the previous two, except in a configuration optimized for EDM measurement precision.  In this case the
horizontal focusing is very ``tame'' but, for optimal EDM tune measurement, the vertical tune $Q_y$ has to 
be tuned toward zero. With $\beta_x$ so nearly constant, the vertical tune is accurately given by 
$Q_y=(2\pi)^{-1}\mathcal{C}/<\beta_x>$, where $\mathcal{C}$ is the ring circumference.}
\end{figure}

A scale-independent, BSM ring beta function plot obtained with the MAPLE-BSM program is shown in 
FIG~\ref{fig:labeled-beta-functions-NOMINAL-map-index}, for the BSM ring layout shown in 
FIG~\ref{fig:PTR-layout-0p80-COSY-4PAX}.  Corresponding, scale-specific results are shown in 
figures~\ref{fig:labeled-beta-functions-NOMINAL-s}, and \ref{fig:labeled-beta-functions-EDM-EXPERIMENT-s}. 
The ring has super-periodicity $n_c=4$. Optics for one super-period is shown on the left in
FIG~\ref{fig:PTR-layout-0p80-COSY-4PAX}. Since each quadrant has element reversal
symmetry, it is sufficient to design, and display, just one eightth of the
ring. $\beta_x$ and $\beta_y$ are plotted against element indices (ordinal integers, starting at 1) in 
FIG~\ref{fig:labeled-beta-functions-NOMINAL-map-index}. (Barely visible) grid lines mark the boundaries
between adjacent elements. One sees that the ring is very simple since the element index increases by 1 
from element to element, and the figure is mirror symmetric about 10. All element names, including
drift lengths, are shown. 

(To preserve site neutrality) the design is length scale invariant. But the length scale for the
similar plot is fixed in FIG~\ref{fig:labeled-beta-functions-NOMINAL-s} and the horizontal axis
is correspondingly changed to longitudinal position $s$ in this and subsequent figures.
(Obviously) the curve shape is distorted, but the vertcal coordinates at the plotted points
are unchanged.  Horizontal and vertical tune advances are plotted on the right. (Note that the left
figure shows one quadrant, while the right one shows only an octant.) The length scale has been
selected such that the ring circumference is about 160\,m; as already mentioned, this can be 
changed with no essential effect to the design..

In a so-called ``NOMINAL'' operational mode shown in FIG~\ref{fig:labeled-beta-functions-NOMINAL-s} ,
the full-ring tunes $Q_x$ and  $Q_y$ are roughly equal, about 3.5.
Most of the planned tests using the ring as an EDM prototype will be done in this configuration. 
This mode of operation is as robust as possible, as regards storage ring operational performance.
Furthermore, in this configuration any influence of electrode shape focusing on the ring optics will 
be negligible and the electrode shapes can be treated as purely cylindrical, $m=0$, enen though they
alternate between $\pm m_{in}$, where $m_{in}$ has a value, not yet fixed, but small compred to 0.1.
With both tunes large compared to 1, the optical behaviour of the lattice will be essentially 
the same as if the bending were magnetic. Only in an EDM-EXPERIMENT mode discussed below, will the lumped 
quadrupole strengths be weak enough for the electrode focusing to be more nearly dominant.
In this limit the ring optics deviates markedly from magnetic ring optics, because of the extra focusing
provided by position dependent electric potential.

Lattice optics for an ``EDM-EXPERIMENT'' mode of operation is shown in
FIG.~\ref{fig:labeled-beta-functions-EDM-EXPERIMENT-s}.  The fundamental phenomenon limiting the 
precision with which the proton (or any other) particle EDM can be measured is ``out of plane''
spin precession induced by unknown radial magnetic field acting on particle MDMs.  Here ``out of plane'' 
means out of the horizontal plane containing the design, central particle, closed orbit.  
(Not counting subsequent averaging over symmetrically varied configurations) the most effective method for suppressing 
this EDM-mimicking precession, 
is to suppress the average vertical separation of counter-circulating beams---like the spurious MDM induced precession, 
this vertical beam separation is proportional to the $\langle B_r\rangle$ average. Suppression of this ``background'' EDM error 
can be described as the storage ring providing ``self-magnetometry''.  For the self-magnetometry to be 
most precise requires the vertical focusing to be weakest possible. As stated in the figure caption, 
with $\beta_y$ so nearly constant, the vertical tune is accurately given by 
$Q_y=(2\pi)^{-1}\mathcal{C}/<\beta_x>$, where $\mathcal{C}$
is the ring circumference. For ultimate EDM accuracy $Q_y$ is expected to be close to zero 
as possible---for example $Q_y=0.01$.  

One sees that ultimate EDM precision will likely require
$\langle\beta_y\rangle$ values an order of magnitude larger than the, already large, value shown in
FIG.~\ref{fig:labeled-beta-functions-EDM-EXPERIMENT-s}. Tuning the ring lattice to achieve this 
by adjusting the BSM lumped quadrupole strengths will be easy; but  preserving 
counter-circulating beams \emph{if possible at all, will not be easy}.  Short of subsequent averaging over
equivaent configurations, this consideration is expected to set the ultimate
achievable EDM precision achievable with this strategy for minimizing $\langle B_r\rangle$.  

The doubly-magic proton-helion measurement, labelled (q1) and (q2) in Table~\ref{tbl:Examples.1},
by measuring the difference of proton or helion EDM's indirectly cancels this source of systematic error,
obviating the need for such extreme rejection of $\langle B_r\rangle$.  So, for the doubly-magic measurement,
the very robust NOMINAL mode of BSM operation may be sufficient.

\subsection{Beam bunching preservation by a single RF accelerating cavity} 
With the two beams having different momenta, their velocities also differ. For both beams to be
captured by the same RF cavity, their harmonic numbers have to differ. The column in 
Table~\ref{tbl:Examples.1} labelled ``best RF harmonic ratio'' gives the harmonic number ratio
best matching the velocity ratio of the two beams, consistent with being small enough for the
RF frequency to be not too large.  Typical radial discrepancies range from very small,
ten's of microns values, almost up to one millimeter. This is taken to be acceptably good 
matching. 

\subsection{Wien filter spin-tune adjustment}
Superimposed electric and magnetic bending fields allow 
small correlated changes of $E$ and $B$ to alter the spin tune 
without affecting the orbit. Being uniformly-distributed, 
appropriately matched electric and magnetic field components 
added to pre-existing bend fields can act as a (mono-directional)
``global Wien filter'' that adjusts the spin tune without changing 
the closed orbit. Replacing the requirement that $\eta_E$ and 
$\eta_M$  sum to 1, we require $\Delta\eta_M=-\Delta\eta_E$, 
and obtain, using the same fractional bend formalism,
for a Wien filter of length $L_W$ the spin tune shift caused by a 
Wien filter of length-strength product  $EL_W$ is given by 
\begin{equation}
\Delta Q_S^W
 =
-\frac{1}{2\pi}\,\frac{1+G}{\beta^2\gamma^2}\,\frac{EL_W}{mc^2/e}.
\label{eq:BendFrac.6pp} 
\end{equation}
For ``global'' Wien filter action by the bends of the entire ring, 
$L_W$ is to be replaced by $2\pi r_0$.  Note though that, even in this case,
the Wien filter produces the pure tune shift given by Eq.~(\ref{eq:BendFrac.6pp})
only for one of two counter-circulating beams; presumeably the primary beam.
The secondary beam closed orbit will vary as the primary beam tune is adjusted,
or stabilized. 

\subsection{``MDM comparator trap'' operation}
This section digresses temporarily to describe
the functioning of dual beams in the same ring as a ``spin tune 
comparator trap''.
A ``trap'' is usually visualized as a ``table-top apparatus''. 
For this paper ``table-top radii'' of $10$, $20$, or $50$, meters
(or rather curved sectors of these radii, expanded by straight sections
of comparable length) are considered.

As mentioned previously, the electron MDM has been determined 
with 13 decimal point accuracy.  Though other magnetic moments are also known to high
accuracy, compared to the electron their accuracies are inferior by 
three orders of magnitude or more. One purpose for a
spin-tune-comparator trap would be to ``transfer'' some of the electron's 
precision to the measurement of other magnetic dipole moments (MDM's). 
For example, the proton's MDM could perhaps be determined to almost the 
current accuracy of the electron's.

Different (but not necessarily disjoint) co- or counter-circulating 
beam categories include different particle type, opposite sign, dual speed, 
and nearly pure-electric or pure-magnetic bending.
Cases in which the bending is nearly pure-electric are easily 
visualized. The magnetic bending ingredient can be treated perturbatively. 
This is especially practical for the 14.5\,MeV electron-electron 
and the 233\,Mev proton-proton counter-circulating combinations.

Eversmann et al.\cite{Eversmann} have demonstrated the capability
of measuring spin tunes with high accuracy. By measuring the spin
tunes of beams circulating in the same ring (preferably, but not necessarily 
simultaneously) the MDM's of the two beams can be accurately compared. 

\section{Doubly-frozen spin EDM measurement examples and methods}
\subsection{Major EDM developments from the past}
Important EDM advances that have been made in past can be listed: 
The storage ring ``frozen spin concept'' according to which, for a 
given particle type, there can be a kinetic energy for which the 
beam spins are ``frozen'' in a storage ring---for example 
always pointing along the line of flight, Farley et al.\cite{Farley};
The recognition of all-electric rings with ``magic'' frozen spin kinetic 
energies (14.5\,MeV for electrons, 233\,MeV for protons) as especially 
appropriate for EDM measurement, Semertzidis et al.\cite{BNL-2011}.
The ``Koop spin wheel'' mechanism in which a small radial magnetic field 
$B_r$ applied to an otherwise frozen spin beam causes the beam polarization 
to ``roll'' around a locally-radial axis\cite{Koop-SpinWheel}, 
(systematic precession around any axis other than this would cancel any 
accumulating EDM effect). Koop\cite{Koop-different-particles} 
has also suggested simultaneous circulation of different particle types, 
though not with the detailed lattice design nor the doubly-frozen spin frequency-domain comagnetometry 
averaging analysed in the present paper.  Spin coherence times long enough for accumulted EDM-induced 
precession to be measureably large has been demonstrated by Eversmann et al.\cite{Eversmann}; 
``Phase-locking'' the beam polarization, which allows the beam polarization 
to be precisely manipulated externally, has been demonstrated by Hempelmann et al.\cite{Hempelmann}.  

\subsection{Cancelation of unknown radial magnetic field $\langle\Delta B_r\rangle$\label{sec:RaialMagField}}
By design, the only intentionally non-zero field components in the proposed ring would be the radial 
electric component $E_x$, and ideally-superimposed magnetic bending would be provided 
by a vertical magnetic field component $B_y$. Routine initial cancelation of $\langle\Delta B_r\rangle$ can be performed using 
unpolarized counter-circulating beams by measuring the  differential vertical separation of 
the two beams, which is similarly proportional to $\langle B_x\rangle$.  

Since the dominant systematic EDM measurement error is proportional to $\langle\Delta B_r\rangle$, in principle
this cancelation is all that is needed to eliminate the dominant systematic error. But the effectiveness of this
cancelation depends on vertical position sensitivity of the beam position monitors (BPMs) and on the 
restoring force of the lattice focusing.  As illustrated in FIG~\ref{fig:labeled-beta-functions-EDM-EXPERIMENT-s},  
this ``self-magnetometer sensitivity'' can be increased only until beam lifetime reduction due to the 
vertical particle loss becomes unacceptably large.

\subsection{Koop spin wheel EDM determination}
By design, the only field components in the proposed ring would be the radial 
electric component $E_x$, and ideally-superimposed magnetic bending would be provided 
by a vertical magnetic field component $B_y$. There also needs to be a tuneable 
radial magnetic field $B_r\equiv B_x$, to compensate any uninentional and 
unknown radial magnetic field and to control the roll-rate of the Koop spin
wheel.

For a ``Koop spin wheel'' rolling around the radial $x$-axis, notes 
by I. Koop\cite{Koop-SpinWheel} provide formulas for the
roll frequencies (expressed here in SI units, with $B\rho$ in T.m),  
\begin{equation}
\Omega_x^{\rm B_x} = -\frac{1}{B\rho}\,\frac{1+G}{\gamma}\,cB_x,
\quad\hbox{and}\quad
\Omega_x^{\rm EDM} =  -\eta\frac{1}{B\rho}\,\bigg(\frac{E_x}{c} + \beta B_y\bigg).
\label{eq:roll-rates} 
\end{equation}
$\Omega_x^{\rm EDM}$ is the foreground, EDM-induced, out-of-plane precession frequency. 
$\Omega_x^{\rm B_x}$ is a roll frequency around the same radial axis,
induced by a radially magnetic field $B_x$ acting on the MDM. 
$cB\rho=pc/(qe)\equiv pc/(Ze)$ is the standard accelerator 
physics specification of storage ring momentum. The factor $\eta$ expresses 
the electric dipole moment $d=\eta\mu$ in terms of the magnetic moment 
$\mu$ of the beam particles.

FIG~7.28 in CERN EDM Feasability Report\cite{CYR} illustrates a ``calibration mode'' in which the
linear dependence of $\Omega_x^{\rm B_x}$ on $B_x$ is determined with high 
precision using the first of Eqs.~(\ref{eq:roll-rates}) and a ``measurement mode'' by which $\eta$
is determined using the second of Eqs.~(\ref{eq:roll-rates}).

Meanwhile, the secondary beam is locked to an unambiguous frequency, depending
only on the $cB_0$ and $E_0$ values.  
Like $B_x$, these bending fields can therefore be set, reversed, and reset to high 
accuracy, based purely on RF and precession frequencies measurements. This resettabilty is expected to 
permit the calibration mode determinations to be performed with high ``frequency domain'' precision. 
These procedures are expected to reduce the systematic EDM error by 2 or 3 orders of magnitude 
beyond that established by the self-magnetometry described in Section~\ref{sec:RaialMagField}
along with the self-magnetometry implied by FIG~\ref{fig:labeled-beta-functions-EDM-EXPERIMENT-s}.

\subsection{Some practical configurations}
Kinematic parameters for some practical doubly-frozen configurations are listed in Table~\ref{tbl:Examples.1}.
Bend radius $r_0$ could be increased beneficially, except for cost, in all cases, \emph{but not necessarily decreased}.
The nominal all-electric, frozen spin proton case, shown in the top row, assumes $r_0=50$\,m.   This futuristic, large 
and expensive, 232.8\,MeV frozen spin proton ring has been referred to as the ``Holy Grail'' facility.
The remaining entries assume radius $R_0=12\,m$, consistent with inexpensive, almost immediate application,
in the COSY, Juelich beam hall.  Proton and deuteron examples are given in a companion paper, presently in preparation.

Master beam (columns on the left) spin tunes are always exactly zero. Spin tunes of secondary beams are given in the final column. 
In all cases they have been calculated closely enough to guarantee they can be tuned exactly to zero.
``Harmonic ratio'' entries indicate optimal RF harmonic number ratios for matching the circumferences of the CW and CCW orbits.  
The fact that these circumferences are not quite equal, wil require the EDM measurements to be corrected accordingly.   

\begin{table}[htb]\footnotesize
\centering
\begin{tabular}{|c||c|c|c|c|c|c|c|c||c|c|c|c|c|c|} \hline 
label &   r0  & CW   & best RF  & QS1  &       KE      &    E0     &   B0     & $\eta_E$ & CCW  & best RF  &    KE2       &  pc2      &  QS2   \\
      &       & beam & harmonic &      &  MeV          &   MV/m    &    T     &          & beam & harmonic &    MeV       &  GeV      &       \\  
      &       &      &  ratio   &      &               &           &          &          &      &  ratio   &              &           &       \\ \hline 
\multicolumn{14}{|c|}{ PERTURBED FROZEN SPIN PROTON-PROTON (nominal all-electric, optional magnetic scanning)} \\ 
 (b)  &  50  &  p    &   1/1   &  0/1  & $\wt{232.8}$  &  8.386   & 1.6e-08   &   1.0    &   p  &   1/1    & $\wt{232.8}$   & -0.7007  & 0/1   \\  \hline
\multicolumn{14}{|c|}{ FROZEN SPIN PROTON-POSITRON (best ultimate proton EDM precision)}           \\  
 (c1) &  12  &  p    &  33/115  &  0/1 & $\wt{86.63}$  & 10.592   &   0.0268  &  0.766   &   e+ &  82/115  & $\wt{30.09}$  & -0.0306  & 0/1    \\ 
\multicolumn{14}{|c|}{ FROZEN SPIN POSITRON-PROTON (inverse of (c1))}                               \\  
 (c2) &  12  &  e+   &  82/115  &  0/1 & $\wt{30.09}$  & 10.592   &  -0.0268  &   4.155  &   p  & 33/115   & $\wt{86.64}$  & -0.4124  & 0/1    \\  \hline 
\multicolumn{14}{|c|}{ FROZEN SPIN HELION-PROTON (determines proton-helion EDM difference)}    \\  
 (q1) &  12  & h     &  85/228  &  0/1 & $\wt{39.24}$  &   4.387  &  -0.0230  &   1.351  &   p  & 143/228   & $\wt{38.59}$ & -0.2719  & 0/1   \\  
\multicolumn{14}{|c|}{ FROZEN SPIN PROTON-HELION (inverse of (q1)) }                               \\  
 (q2) &  12  & p     & 143/228  &  0/1 & $\wt{38.59}$  &   4.387  &   0.0230  &  0.6958  &   h  & 85/228    & $\wt{39.24}$ & -0.4711  & 0/1   \\  \hline 
\end{tabular}
\caption{\label{tbl:Examples.1}
Sample beam-pair combinations for the EDM experiments discussed in this paper; master beam entries 
on the left, secondary beam on the right. ``(b)'', ``(c1)'', etc. are case labels, copied from a previous report\cite{Doubly-frozen}. 
Dual rows allow either particle type to be designated ``primary beam''.  Overhead tildes $\widetilde{\ \ }$, indicate values known to much greater
accuracy, but truncated for display in this table.  Candidate beam particle types are ''e+'',``p'', ``d'', ``t'', ``h'' that could label 
label positron, proton, deuteron, triton, or helion rows. Proton and deuteron examples are given in a companion paper. Bend radii, particle type, 
and kinetic energies are given in the first three columns. There is no fundamental dependence of spin tune $Q_s$ on $r_0$, but $r_0$ values have 
been chosen to limit $|E_0|$ to realistic values.  All but the top entry assume bend radius $r_0=12$\,m, but the required electric field $E_0$
may be unrealistically large in some cases.}
\end{table}

\subsection{Estimation of MDM and EDM measurement precisions}
The ``dipole moment comparator'' name proposed for the class of storge rings described in this paper intentionally applies to both magnetic and electric dipole moments.  Strictly speaking, since the dimensionalities of these quantities are different, for them to be comensurate requires a qualification defining comparably strong electric and magnetic field values, such as E=cB in MKS units.  Even with this qualification, because parity and time reversal symmetries suppress EDMs so strongly, it is not appropriate to compare the fractional accuracies of MDMs and EDMs.  It is more nearly appropriate to compare the absolute precisions of MDM and EDM measurement.

Once this limitation is accepted, it becomes sensible to concentrate on the precision with which EDMs can be measured---any measurably non-zero EDM value would imply a measurement error which---applied to any MDM (except the electron's\cite{Gabrielse-eEDM})---would represent a fractional MDM determination smaller than current limits.  For brevity then, it is sufficient to discuss only the precision with which elementary particle EDMs can be measured.

The top entry in Table~\ref{tbl:Examples.1} applies to any propoosed proton EDM measurement in which 232.8\,MeV frozen spin proton beams counter-circulate simultaneously in a ring with all-electric bending, as proposed, for example, in references\cite{BNL-2011},\cite{Anastassopoulos}, or \cite{hybrid}.
The achievable proton EDM systematic error in these papers is said to not exceed $10^{-29}$\,e-cm.  It is the unknown maximum average radial magnetic fieild that establishes this limit in the first two cases.  An independent re-analysis of this class by Valeri Lebedev\cite{Lebedev} stated that ``it is not feasible for the average radial magnetic field to be suppressed below 1\,nG---below the assumed value by about 4 orders of magnitude.''  The PTR ring displayed in Figure~\ref{fig:PTR-layout-0p80-COSY-4PAX} has been proposed\cite{CYR} as a prototype for the all-electric 232.8\,MeV proton ring\cite{Anastassopoulos} as well ``hybrid'' rings with magnetic focusing\cite{hybrid}.   

An all-electric ring of our EDM comparator design applied to the proto/proton case measures the difference of the proton EDM with itself---which is, of course, zero; this will provide a useful consistency test.  When applied to the proton/helion case, it is the difference of proton and helion EDM's that is measured.   What is special about this case is that the dominant systematic error cancels, leaving a statistical error limit of about $10^{-30}$\,e-cm as dominant error.  Since the difference of vanishingly small quantities is vanishingly small, any measurably large result would provide evidence of physics beyond the standard model.  

To achieve such a small statistical error will require averaging runs with proton and helion beams interchanged.  The precision with which magnetic field reversal can be achieved with the required precision is controlled digitally by simultaneously phase locking the spin tunes of both simultaneously counter-circulating beams.  This strategem exploits the particle magnetic dipoles as perfect stabilizing gyroscopes for the establishment, stabilization, reproducibility and field reversability of \emph{in-plane precession} to enable the measurement of \emph{out-of-plane precession} induced by any non-vanishing EDMs.  Under the near-certain assumption that the positron EDM is negligibly small, the proton/positron entry in Table~\ref{tbl:Examples.1} will provide a direct measurement of the proton EDM at the same  $10^{-30}$\,e-cm accuracy level.

\acknowledgments
This paper has profited from collaboration with colleagues at the COSY laboratory
in Juelich, Germany; especially Sig Martin (co-ring-designer, now sadly deceased), Kolya Nikolaev, Frank Rathmann, 
and Hans Stroeher (especially in connection with the present paper)
and Ralf Gebel, Alexander Nass and Helmut Soltner (responsible for matching electric and magnetic profiles)
and Andreas Lehrach (who pointed out an error in the symmetry of the 
original electrode focusing arrangement in FIG~\ref{fig:PTR-layout-0p80-COSY-4PAX}). 
Colin Wilkin provided important background material concerning polarized beam scattering capabilities at COSY.
Acknowledgement for help and discussions is due also 
to my colleagues at Cornell, espcially Eanna Flanagan, Lawrence Gibbons, Maxim Perlstein, Saul Teukolsky,  
and Anders Ryd. Also to Lindsay Schachinger, Nikolay Malitsky, and to my son John Talman for essential 
TEAPOT, UAL, and ETEAPOT code development.  Also to Bill Marciano, Bill Morse, Steve Peggs and Yannis 
Semertzidis at BNL, to Yunhai Cai, Alex Chao, Michael Peskin, John Seeman and Gennady Stupakov at SLAC, 
to Christan Carli (who pointed out the impracticality of pure electrode focusing, and contributed to this paper 
in various other ways), Mike Lamont, and Malik Tahar (who, with Carli, helped with injection design) at CERN, 
as well as Joe Grames and Matt Poelker at Jefferson Lab, all for valuable discussions.

\end{document}